# A 'New Formula' to Provide the Compatibility between the Special Theory of Relativity (STR), Black Holes and Strings

**Ali Riza AKCAY**

TUBITAK-UEKAE

P.K. 21, 41470 - Gebze

Kocaeli-TURKEY

E-mail: aakcay@yunus.mam.gov.tr

**Abstract**

This paper describes a 'New Formula' in place of Einstein's Famous Formula (EFF) to provide the compatibility between the Special Theory of Relativity (STR), black holes and strings. The 'New Formula' can also predict and describes the space-time singularities without the distribution of mass and energy. According to the 'New Formula', any particle can reach to and exceed the speed of light $(v \geq c)$.

The EFF $(E = mc^2)$ is only valid and applicable in the vacuum (the mediums which have low current density: outside the string, outside black hole), but is not valid and applicable for inside string and inside black hole including space-time singularities. However, the 'New Formula' is valid and applicable in all mediums including inside string and inside black hole.

## 1. Introduction

As known, the Einstein's Famous Formula $\left[E = m_0 c^2 / \sqrt{1 - \boldsymbol{b}^2}\right]$ is the general result of the Special Theory of Relativity (STR). According to this formula; the energy ($E$) approaches infinity as the velocity $v$ approaches the velocity of light ($c$). The velocity (or speed) must therefore always remain less than $c$. This means that it is not allowed by the STR to travel faster than light [1].

A black hole is a region of space from which it is impossible to escape if one is traveling at less than the speed of light. But the Feynman sum over histories says that particles can take any path through space-time. Thus it is possible for a particle to travel faster than light [2].

What is meant was that matter could curve a region in on itself so much that it would effectively cut itself off from the rest of the Universe. The region would become what is called a black hole. Objects could fall into the black hole, but noting could escape. To get out, they would need to travel faster than the speed of light, which is not allowed by the theory of relativity. Thus the matter inside the black hole would be trapped and would collapse to some unknown state of very high density. Einstein was deeply disturbed by the implications of this collapse, and he refused to believe that it happened. But, Robert Oppenheimer showed in 1939 that an old star of more than twice the mass the sun would inevitably collapse when it had exhausted all its nuclear fuel [2].

The fact that Einstein's general theory of relativity turned out to predict singularities led to a crisis in physics. The equations of general relativity, which relate to curvature of space-time with the distribution of mass and energy, cannot be defined as a singularity. This means that general relativity cannot predict what comes out of a singularity. In particular, general relativity cannot predict how the universe should begin at the big bang. Thus, general relativity is not a complete theory. It needs an added ingredient in order to determine how the universe should begin and what should happen when matter collapses under its own gravity [2].

Stars which collapse into black holes generally posses a magnetic field. In addition, black holes swallow electrically charged particles from the interstellar medium such as electrons and protons. It is therefore reasonable to expect black holes to have electromagnetic properties. H. Reissner in 1916, and independently G. Nordstrom in 1918, discovered an exact solution to Einstein's equations for the gravitational field caused by an electrically charged mass. This solution is generalized version of Schwarzschild's solution, with one other parameter: the electric charge. It describes space-time outside the event horizon of an electrically charged black hole [3]. The 'New Formula' can describe the electrically charged black holes by using the current density as a parameter.

The most remarkable discovery including a semi-classical gravitational effect is the Hawking radiation, which is concluded by treating matter fields on space-time as quantum while a black hole metric as classical. According to this theory, a black hole radiates particle flux of a thermal spectrum, whose temperature is $k/2p$ where $k$ is the surface gravity [4].

The existence of Hawking radiation is closely related to the fact that a particle which marginally escapes from collapsing into a black hole is suffered from infinite redshift. In other words, the particle observed in the future infinity had a very high frequency when it was near the event horizon.

Gamma-ray bursts (GRBs) appear as the brightest transient phenomena in the Universe. The nature of the central engine in GRBs is a missing link in the theory of fireballs to their stellar mass progenitors. It is shown that rotating black holes produce electron-positron outflow when brought into contact with a strong magnetic field. The outflow is produced by a coupling of the spin of the black hole to the orbit of the particles. For a nearly extreme Kerr black hole, particle outflow from an initial state of electrostatic equilibrium has a normalized isotropic emission of $\sim 5\times 10^{48}(B/B_c)^2(M/7M_O)^2 \sin^2 q$ erg/s, where $B$ is the external magnetic field strength, $B_c = 4.4\times 10^{13} G$, and $M$ is the mass of the black hole. This initial outflow has a half-opening angle $q \geq \sqrt{B_c/3B}$. A connection with fireballs in $g$-ray bursts is given [12].

Cosmic strings can be turned into superconductors if electromagnetic gauge invariance is broken inside the strings. This can occur, for example, when a charged scalar field develops a non-zero expectation value in the vicinity of the string core. The electromagnetic properties of such strings are very similar to those of thin superconducting wires, but they are different from the properties of bulk superconductors [11].

Strings predicted in a wide class of elementary particle theories behave like superconducting wires. Such strings can carry large electric currents and their interactions with cosmic plasmas can give rise to a variety of astrophysical effects [11].

To provide the compatibility between STR, black holes and strings a 'New Formula' has been developed by adding the ratio $(J/J_{max})$ as a new parameter (or dimension) to the EFF.

## 2. Superconductivity

The discovery of superconductivity started from the finding of Kamerlingh Onnas in 1911 that the resistance of mercury has an abrupt drop at a temperature of 4.2 $^0$K and has practically a zero dc-resistance value at temperatures below 4.2 $^0K$. This new phenomenon of zero-resistance at low temperature was soon found in many other metals and alloys. An important characteristic of the loss of dc-resistance observed is the sharpness of the transition. The temperature at which superconductivity first occurs in a material is thus termed the critical (or transition) temperature of the material and is denoted by $T_c$ [5].

A superconductor is simply a material in which electromagnetic gauge invariance is spontaneously broken. Detailed dynamical theories are needed to explain why and at what temperatures this symmetry breaking occurs, but they are not needed to drive the most striking aspects of superconductivity: exlusion of magnetic fields, flux quantization, zero resistivity, and alternating currents at a gap between superconductors held at different voltages [6].

## 2.1 The discovery of High-$T_c$ Superconductors

The first of a new family of superconductors, now usually known as the *High-$T_c$* or *cuprate* superconductors, was discovered in 1986 by Bednorz and Müller. It was a calcium-doped lanthanum cuprate perovskite. When optimally doped to give the highest $T_c$, it had the formula $La_{1.85}Ca_{0.15}CuO_4$, with a $T_c$ of 30 $^0K$. This was already sufficiently high to suggest to the superconductivity community that it might be difficult to explain using the usual forms of BCS theory, and a large number of related discoveries followed quickly. In the following year Wu *et al* found that the closly related material $Yba_2Cu_3O_{7-\delta}$, now known as YBCO, has a $T_c$ of about 93 $^0K$ when $\delta \cong 0.10$, well above the boiling point of liquid nitrogen [7].

## 2.2 Basic Superconductivity: The Order Parameter

A relativistic version of a superconductor the abelian Higgs model

$$L = -\frac{1}{4} F_{mn} F_{mn} + (D_m \Phi)^* (D_m \Phi) - V(\Phi) \tag{1}$$

$F_{mn}$ is the electromagnetic field strength, $D_m$ is the covariant derivative

$$D_m \Phi = (\partial_m - iqA_m) \Phi \tag{2}$$

and $V(\Phi)$ the potential of the scalar field

$$V(\Phi) = \frac{1}{4} (\Phi^* \Phi - m^2)^2 \tag{3}$$

If $m^2$ is positive the field $\Phi$ has a nonzero vacuum expectation value.

A convenient parametrization of $\Phi$ is

$$\Phi = r e^{iqq} \qquad r = r^* > 0 \tag{4}$$

Under gauge transformations

$$A_m \to A_m - \partial_m a \qquad q \to q + a \tag{5}$$

The covariant derivative (2) reads in this notation

$$D_m \Phi = e^{iqq} [\partial_m - iq(A_m - \partial_m q)] r \tag{6}$$

The quantity $\tilde{A}_m = A_m - \partial_m q$ is gauge invariant. Moreover

$$F_{mn} = \partial_m A_n - \partial_n A_m = \partial_m \tilde{A}_n - \partial_n \tilde{A}_m \tag{7}$$

The equation of motion reads, neglecting loop corrections (or looking $L$ as an effective lagrangian)

$$\partial_m F^{mn} + \frac{\tilde{m}^2}{2} \tilde{A}_n = 0 \qquad \tilde{m} = \sqrt{2} q \langle \Phi \rangle \tag{8}$$

In the gauge $A_0 = 0$ a static configuration has $\partial_0 \vec{A} = 0$, $\partial_0 \Phi = 0$ so that $E_i = F_{0i} = 0$. Eq.(8) implies that

$$\vec{\nabla} \wedge \vec{H} + \frac{\tilde{m}^2}{2} \vec{\tilde{A}} = 0 \qquad (9)$$

The term $\tilde{m}^2/2 \vec{\tilde{A}}$ in Eq.(9) is a consequence of spontaneous symmetry breaking and is an stationary electric current (London current). A persistent current with $\vec{E} = 0$, means $\mathbf{r} = 0$ since $\mathbf{r}\vec{j} = \vec{E}$ and hence superconductivity. The curl of Eq. (9), reads

$$\nabla^2 \vec{H} - \frac{\tilde{m}^2}{2} \vec{H} = 0 \qquad (10)$$

The magnetic field has a finite penetration depth $1/\tilde{m}$, and this nothing but the Meissner effect. The key parameter is $\langle \Phi \rangle$, which is the order parameter for superconductivity: it signals spontaneous breaking of charge conservation [8].

## 2.3 Critical current density

The following set of differential equations are called Ginzburg-Landau equations:

$$\mathbf{a}\mathbf{y} + \mathbf{b}|\mathbf{y}|^2 \mathbf{y} + \frac{1}{2m^*}\left(-i\hbar \nabla - e^* A\right)^2 \mathbf{y} = 0 \qquad \text{in V} \qquad (11)$$

$$J = \frac{e^* \hbar}{i 2 m^*}\left(\mathbf{y}^* \nabla \mathbf{y} - \mathbf{y} \nabla \mathbf{y}^*\right) - \frac{e^{*2}}{m^*}|\mathbf{y}|^2 A \qquad \text{in V} \qquad (12)$$

We have written $J$ for $J_s$ (supercurrent density) since in thermodynamic equilibrium there are no normal currents.

By noting $\mathbf{y} = |\mathbf{y}|\exp(i\mathbf{q})$, equation for the supercurrent density may also be written as

$$J = \frac{e^*|\mathbf{y}|^2 \hbar}{m^*}\left(\nabla \mathbf{q} - \frac{e^*}{\hbar} A\right) \qquad (13)$$

which shows that the gradient of the phase of the wave function $\mathbf{y}$ determines the observable quantity, the supercurrent density [5].

We shall now apply G-L (Ginzburg-Landau) equations to calculate the critical current in a superconducting thin film at which superconductivity breaks down. First, we consider thin film $d \ll x$ and $d \ll l$ so that $|y|$ and $J_s$ may be supported to be a constant and uniform over the sample cross section of the thin films. We can set $y = |y|e^{iq(x)}$ with $|y|$ being independent of $x$. Equation (13) for the supercurrent yields

$$J = \frac{e^*}{m^*}\left(\hbar \nabla q - e^* A\right)|y|^2 = e^*|y|^2 \mathbf{n}_s \tag{14}$$

where $\mathbf{n}_s$ denotes the mean velocity of the superconducting pair of electrons [5].

*The mixed-state critical current density as a function of B*: The critical current density $J_c$ is the transport current density at which pinning can no longer hold the flux at rest in the face of the thermodynamic driving force. At this point the frictional pinning force per unit volume $f_f / l^3$ is equal to $J_c \Lambda B$, so we have

$$J_c \cong \frac{1}{l^3 B}\left[\mathbf{zh}(1 - B/B_{c2})(B_c^2/2\mathbf{m}_0)p^2\right]. \tag{15}$$

It is interesting to write down the ratio of critical current density predicted by (15) to the ideal critical current density $J_{max}$ of a thin film at which superconducting state itself collapses [5].

$$\frac{J_c}{J_{max}} \cong \mathbf{zh}(1 - B/B_{c2})\frac{B_c}{B}\frac{p^2 \mathbf{l}}{l^3}. \tag{16}$$

## 2.4 High-frequency conductivity

Unfortunately, in considering the response of superconductors to high-frequency fields, there are many situations, especially with conventional superconductors, where the non-locality of the response is important, and calculations must be based on the full Mattis-Bardeen equation.

In *dirty limit* (where $\ell \ll d$ and $\ell \ll x_0$) the relative shortness of $\ell$ means that we may replace $I(w, R, T)$ by $I(w, 0, T)$. Assuming that the dominant scattering is elastic scattering by impurities we know that $\ell$ is the same in the normal and superconducting states. We then have an effective complex conductivity $s$ usually written as $s_1 - is_2$ and given by

$$\frac{s_1 - is_2}{s_n} = \frac{I(w, 0, T)}{-ip\hbar w}. \qquad (17)$$

The real part of the Mattis-Bardeen conductivity $s_1(T)$ corresponds to a current of normal excitations. At low temperatures it is exponentially small at low frequencies, varying as $e^{-\Delta/kT}$, but it rises rapidly as soon as $w$ exceeds the *gap frequency* $w_g = 2\Delta/\hbar$ (of order $10^{11}$ to $10^{12}$ Hz for conventional superconductors and $10^{13}$ Hz in cuprates), the frequency at which creation of pairs of excitations becomes possible. The imaginary part of the conductivity $s_2(T)$, corresponding to the superelectrons, is proportional to $1/w$, as one would expect for an inertia-dominated response, almost up to gap frequency. In fact, if we use the Pippard equation as an approximation we find that

$$s_2 \cong \frac{\ell}{x_0} \frac{1}{w\Lambda} = s_n \frac{p\Delta_0}{\hbar w} \frac{\Lambda(0)}{\Lambda(T)}. \qquad (18)$$

This dirty limit conductivity may be compared with the clean limit London conductivity $1/w\Lambda(T)$. Near the gap frequency this approximation fails and $s_2$ falls more rapidly with frequency.

In the context of high-frequency conductivity the two-fluid model means a system whose conductivity may be written as

$$\boldsymbol{s} = \frac{ne^2}{m_e}\left[\frac{f_n}{i\boldsymbol{w}+1/\boldsymbol{t}} + \frac{f_s}{i\boldsymbol{w}+s}\right] \tag{19}$$

where $f_n$ and $f_s$ represent the fractions of the electrons which are normal and superfluid respectively (with $f_n + f_s = 1$), $\boldsymbol{t}$ is a relaxation time for the normal electrons and $s$ is an infinitesimal. In this simple model the normal electrons have both inertia and damping, with the usual Drude conductivity at high frequencies, and the superelectrons have inertia but no dumping . We notice that the model obeys the *conductivity sum rule*

$$\int_{-\infty}^{\infty} \boldsymbol{s}^{'}(\boldsymbol{w})d\boldsymbol{w} = \boldsymbol{p}\frac{ne^2}{m_e} \tag{20}$$

which applies to all systems of mobile electrons [7].

## 2.5 High Current in General Relativity

Physical nature of equilibrium of current-carrying filaments is studied on the basis of Einstein equations of General Relativity. Considering a conducting filament as an element of the structure of universe, one has to take into account both electromagnetic and gravitational interactions of charges [9].

Intergalactic currents are playing a very important role in modern plasma astrophysics. Understanding that the universe is largely a Plasma Universe came from the fact that electromagnetic forces exceed gravitational forces by a factor $10^{36}$, and even if neutral as a whole system a relatively small electromagnetic fluctuation can lead to non-uniform distribution of matter [9].

Another topic that requires General Relativity is the old Alfven's problem of a limiting current. If we ignore the effect of General Relativity, then self-consistent theory does not impose any limitations on the current values of equilibrium relativistic beams. In General Relativity the matter curves the space-time, and this results in gravitational self-attraction of matter. If total energy (or mass) of matter exceeds some limit, the forces of contraction cannot be balanced by the pressure. In this case equilibrium is not possible, and the matter undergoes infinite contraction, which is called gravitational collapse [9].

We show that the current of an equilibrium filament cannot exceed $I_{max} = 0.94 \cdot 10^{25}$ A. Currents $I \approx 10^{20}$ A in the Galactic and Intergalactic Medium are discussed by Peratt. Nevertheless solution of the problem of limiting current in General Relativity is interesting in principle, especially taking into account filamentary structure of the Universe. Our analysis realizes common physical nature of gravitational and electromagnetic collapse, and displays peculiarities of space distribution of matter and gravitational field near the collapse boundary [9].

## 3. Background for the Evolution of the 'New Formula'

This chapter gives some background information that upholds the 'New Formula'.

## 3.1 Acceleration of Ultra High Energy Particles by Black Holes and Strings (Currents in High Energy Astrophysics)

It is well known that charged black holes can have a magnetic dipole moment (indeed for a rotating charged black hole, the gyromagnetic ratio is 2, the same as for a Dirac particle). Such a black hole can thus also interact with a particle having a magnetic moment. The interaction energy in this case is given by:

$$E_{int} \cong \frac{^mBH\,^mP}{r^3} \times curved\ space\ factors \qquad (21)$$

Here $^mBH$ and $^mP$ are the magnetic dipole moments of the black hole and the particle respectively.

For $^mP \approx {}^mB$ (the Bohr magneton) and for maximally charged hole this gives a maximal energy (at $r = r_s$) of:

$$E_{max} \cong \frac{{}^mBc^4}{MG^{3/2}} \tag{22}$$

For a $10^{17}$ gm. primordial Hawking black hole this gives (note $E \alpha 1/M$)

$$E_{max} \cong 10^{23} eV.$$

If the black hole is embedded in a magnetic field such high energy particles accelerated by the hole can also emit ultra high frequency gamma radiation (suppressed by $1/m^4$).

We next consider the acceleration of particles by cosmic strings and fundamental superstrings. Superstrings are produced near the Planck scale (energy $E_{pl}$ or $M_{pl} \approx 10^{19}$ GeV). They are characterized by a tension $T_{pl} \approx c^2/G$ (mass per unit length). $T_{pl} \approx 10^{28} gcm^{-1}$ strings produced by a symmetry breaking at any other energy (mass scale) $\approx M$ have a tension given by:

$$T_s \cong \frac{c^2}{G}\left(\frac{M}{M_{pl}}\right)^2 \tag{23}$$

In addition, one can have conducting cosmic strings which are essentially topological line defects [10].

There are some nice analogies between vortex lines in a Type II superconductor (carrying a quantized flux $\hbar c/2e$) and conducting cosmic strings. For instance, the field vanishes everywhere in a superconductor (Meissner effect), i.e. $F_{ab} = 0$, everywhere except along Abrikosov vortex lines carrying a confined quantized flux $\hbar c/2e$. Inside a superconductor we have the Landau equations:

$$\Delta^2 \vec{B} + \mathbf{l}^2 \vec{B} = \vec{J} \tag{24}$$

The vanishing of the field inside a superconductor is an effect of the Landau-Ginzburg theory where we have the Maxwell field coupled to a scalar field as:

$$D_m D^m \mathbf{f} = \mathbf{a} \mathbf{f} \left( |\mathbf{f}|^2 - \mathbf{l}^2 \right) \tag{25}$$

$|\mathbf{f}| \cong \mathbf{l}$ near the broken symmetric state.

Far from the flux tube:

$$D_m \mathbf{f} = \left( \mathbf{d}_m + ieA_m \right) \mathbf{f} = 0 \tag{26}$$

and

$$\left[ D_m, D_n \right] \mathbf{f} = ieF_{mn} \mathbf{f} = 0 \tag{27}$$

So either $\mathbf{f}$ or $F_{mn}$ must vanish. This has the solution:

$$A_m = -(1/e) \mathbf{d}_m \mathbf{f}, \mathbf{f} = \mathbf{l} e^{iq} \tag{28}$$

The Higgs field responsible for these defects is described by a relativistic version of the Landau-Ginzburg model and consequently it can be shown that conducting strings also carry flux.

$$\mathbf{f} = n\hbar c / e \tag{29}$$

The flux can be shown to give rise to an electric field given by:

$$V \cong cT_s G M_{pl}^2 / e\hbar \cong \frac{Gc}{e\hbar} T_s M_{pl}^2 \tag{30}$$

Thus charged particles can be accelerated to a maximal energy given by: (corresponding to a critical current):

$$E \cong ecT_s^{1/2} G^{1/2} M_{pl} \tag{31}$$

For a string tension, corresponding to a GUT scale $M \cong 10^{15} GeV$, (the corresponding tension being given by eq. (23)):

$$E \cong 10^{21} eV \qquad (32)$$

A higher string tension $T_s$ gives rise to a higher value of $E$. For a GUTs scale $M \approx 10^{16} GeV$, $E \approx 10^{22} eV$.

It must be noted that Ultra High Energy (UHE) particles can be spontaneously generated by Evaporating Black Holes (EBH) [10].

## 3.2 Electron-Positron Outflow from Black Holes

Gamma-ray bursts (GRBs) appear as the brightest transient phenomena in the Universe. The nature of the central engine in GRBs is a missing link in the theory of fireballs to their stellar mass progenitors. It is shown that rotating black holes produce electron-positron outflow when brought into contact with a strong magnetic field. The outflow is produced by a coupling of the spin of the black hole to the orbit of the particles. For a nearly extreme Kerr black hole, particle outflow from an initial state of electrostatic equilibrium has a normalized isotropic emission of $\sim 5 \times 10^{48} (B/B_c)^2 (M/7M_O)^2 \sin^2 q$ erg/s, where $B$ is the external magnetic field strength, $B_c = 4.4 \times 10^{13} G$, and $M$ is the mass of the black hole. This initial outflow has a half-opening angle $q \geq \sqrt{B_c/3B}$. A connection with fireballs in $g$-ray bursts is given [12].

A theory is decribed for electron-positron pair-creation powered by a rapidly spinning black hole when brought into contact with a strong magnetic field. The magnetic field is supplied by the surrounding matter as in forementioned black hole/torus or disk systems. A rapidly spinning black hole couples to the surrounding matter by Maxwell stresses [12].

Pair-creation can be calculated from the evolution of wave-fronts in curved spacetime, which is well-defined between asymptotically flat in- or out-vacua. By this device, any inequivalence between them becomes apparent, and generally gives rise to particle production. It is perhaps best known from the Schwinger process, and in dynamical spacetimes in cosmological scenarios. Such particle production process is driven primarily by the jump in the zero-energy levels of the asymptotic vacua, and to a lesser degree depends on the nature of the transition between them. The energy spectrum of the particles is ordinarily nonthermal, with the notable exception of the thermal spectrum in Hawking radiation from a horizon surface formed in gravitational collapse to a black hole. There are natural choices of the asymptotic vacua in asymptotically flat Minkowski spacetimes, where a time-like Killing vector can be used to select a preferred set of observers. This leaves the in- and out-vacua determined up to Lorentz transformations on the observers and gauge transformations on the wave-function of interest. These ambiguities can be circumvented by making reference to Hilbert spaces on null trajectories – the past and future null infinities $J^{\pm}$ in Hawking's proposal – and by working with gauge covariant frequencies. The latter received some mention in Hawking's original treatise, and is briefly as follows [12].

Hawking radiation derives from tracing wave-fronts from $J^{+}$ to $J^{-}$, past any potential barrier and through the collapsing matter, with subsequent Bogolubov projections on the Hilbert space of radiative states on $J^{-}$. This procedure assumes gauge covariance, by tracing wave-fronts associated with gauge-covariant frequencies in the presence of a background vector potential $A_a$. The generalization to a rotating black hole obtainsby taking these frequencies relative to real, zero-angular momentum observers, whose world-lines are orthogonalto the azimuthal Killing vector as given by $x^a \partial_a = \partial_t - (g_{tf}/g_{ff})\partial_f$. Then $x^a \sim \partial_t$ at infinity and $x^a \partial_a$ assumes corotation upon approaching the horizon, where $g_{ab}$ denotes the Kerr metric. This obtains consistent particle-antiparticle conjugation by complex conjugation among all observers, except for the interpretation of a particle or an antiparticle. Consequently, Hawking emission from the horizon of a rotating black hole gives rise to a flux to infinity

$$\frac{d^2n}{dwdt} = \frac{1}{2p} \frac{\Gamma}{e^{2p(w-V_F)/k}+1},$$

for a particle of energy $w$ at infinity [12]. Here, $k = 1/4M$ and $\Omega_H$ are the surface gravity and angular velocity of the black hole of mass $M$, $\Gamma$ is the relevant absorption factor. The Fermi-level $V_F$ derives from the (normalized) gauge-covariant frequency as observed by a zero-angular momentum observer close to the horizon, namely, $w - V_F = w_{ZAMO} + eV = w - n\Omega_H + eV$ for a particle of charge $-e$ and azimuthal quantum number $n$, where $V$ is the potential of the horizon relative to infinity. The results for antiparticles (as seen at infinity) follow with a change of sign in the charge, which may be seen to be equivalent to the usual transformation rule $w \rightarrow -w$ and $n \rightarrow -n$.

### 3.3 Superconducting Strings

Superconductivity can be understood as a spontaneously broken electromagnetic gauge invariance. When the gauge invariance is broken, the photon acquires a mass and any magnetic field applied at the boundary of the superconductor decays exponentially towards its interior. The magnetic field is screened by a non-dissipative superconducting current flowing along the boundry – the well-known Meiβner effect [11].

Cosmic strings can be turned into superconductors if electromagnetic gauge invariance is broken inside the strings. This can occur, for example, when a charged scalar field develops a non-zero expectation value in the vicinity of the string core. The electromagnetic properties of such strings are very similar to those of thin superconducting wires, but they are different from the properties of bulk superconductors [11].

Strings predicted in a wide class of elementary particle theories behave like superconducting wires. Such strings can carry large electric currents and their interactions with cosmic plasmas can give rise to a variety of astrophysical effects [11].

The idea that strings could become superconducting was first suggested in a pioneering paper by Witten [1985a]. Later it was realized that the role of the superconducting condensate could be played not only by a scalar field, but also by a vector field whose flux is trapped inside a non-abelian string [Preskill, 1985; Everett, 1988]. If the vector field is charged, the gauge invariance is again spontaneously broken inside the string. Witten also proposed another mechanism for string superconductivity, which operates in models where some fermions acquire their masses from a Yukawa coupling to the Higgs field of the string [11].

### 3.3.1 Bosonic string superconductivity

The simplest example of scalar string superconductivity occurs in a toy model with two complex scalar fields $f$ and $s$ interacting with separate $\tilde{U}(1)$ and $U(1)_Q$ gauge fields $\tilde{A}_m$ and $A_m$, respectively [Witten, 1985a]. The first $\tilde{U}(1)$ is broken and gives rise to vortices. The second $U(1)_Q$ which we identify with electromagnetism, although unbroken in vacuum, can provide a charged scalar condansate in the string interior. The Lagrangian is merely a replicated version of the abelian-Higgs model,

$$L = |\tilde{D}_m f|^2 + |D_m s|^2 - V(f,s) - \frac{1}{4}\tilde{F}^{mn}F_{mn} - \frac{1}{4}F^{mn}F_{mn}, \quad (33)$$

where $\tilde{D}_m f = \partial_m f - ig\tilde{A}_m f$ and $D_m s = \partial_m s - ieA_m s$.

From the Lagrangian (33) we can derive the usual electromagnetic current density,

$$j_m = ie(\bar{s}D_m s - s\bar{D}_m \bar{s}) \quad (34)$$

The total current $J$ can than be found by integrating over the string cross-section [11]. Assuming that the vector potential $A_m$ remains approximately constant across the string we have

$$J = 2\sum e(\partial_z q + eA_z), \quad (35)$$

where

$$\sum = \int dxdy |s|^2. \quad (36)$$

Using the expression for the current (35), we obtain the total current flowing around the loop [Witten, 1985a]

$$J = \frac{2\sum e}{1 + (\sum e^2/p)\ln(R/d)} \frac{N}{R}. \quad (37)$$

### 3.3.2 String electrodynamics

The defining property of a superconducting string is its response to an applied electric field: the string develops an electric current which grows in time,

$$dJ/dt \sim (ce^2/\hbar)E. \quad (38)$$

Here, $E$ is the field companent along the string and $e$ is the elementary charge $(e^2 \sim 10^{-2})$.

The charge carries in the string can be bosons or fermions. We consider first the case of fermionic superconductivity. Models of this type have fermions which are massless inside the string and have a finite mass $m$ outside the string. Particles inside the string can be thought of as a one-dimensional Fermi gas. When an electric field is applied, the Fermi momentum grows as $\dot{p}_F = eE$, and the number of fermions per unit length, $n = p_F/2\pi\hbar$, also grows [11]:

$$\dot{n} \sim eE/\hbar. \tag{39}$$

The particles move along the string at the speed of light. The resulting current is $J = enc$, and $dJ/dt$ is given by (38).

The current continues to grow until it reaches a critical value

$$J_c \sim emc^2/\hbar, \tag{40}$$

when $p_F = mc$. At this point, particles at the Fermi level have sufficient energy to leave the string. Consequently, in this simplified picture, the growth of the current terminates at $J_c$ and the string starts producing particles at the rate (39). The fermion mass $m$ is model-dependent, but it does not exceed the symmetry breaking scale of the string, $\eta$. Hence,

$$J_c \leq J_{max} \sim e\left(\eta c^3/\hbar\right)^{1/2}, \tag{41}$$

where we have used the relation $m = \eta^2 c/\hbar$. Grand unification strings can carry enormous currents, $J_{max} \sim 10^{31}\ esu/s$, while for electroweak-scale strings $J_{max} \sim 10^{17}\ esu/s$. Note that the actual value of the critical current $J_c$ is highly model-dependent [11].

Superconducting strings can also have bosonic charge carriers. This occurs when a charged scalar or gauge field develops a vacuum expectation value inside the string. As a result the electromagnetic gauge invariance inside the string is broken, indicating superconductivity. The critical current $J_c$ for this type of string is determined by the energy scale at which the gauge invariance is broken. It is model-dependent, but is still bounded by $J_{max}$ from (41) [11].

Superconducting strings can develop currents not only electric, but also in magnetic fields. Consider a segment of string moving at a speed $v$ in magnetic field $B$. In its rest frame the string 'sees' an electric field $E \sim (v/c)B$, and so the current grows at the rate

$$dJ/dt \sim (e^2/\hbar)vB. \qquad (42)$$

A closed loop of length $L$ oscillating in a magnetic field acts as an ac generator and develops an ac current of amplitude

$$J \sim 0.1(e^2/\hbar)BL \qquad (43)$$

The factor of 0.1 appears because the area of the loop is typically of the order $A \sim 0.1L^2$.

An oscillating current-carrying loop in vacuum emits electromagnetic waves. For a loop without kinks or cusps the radiation power is

$$\dot{E}_{em} \sim \Gamma_{em} J^2/c, \qquad (44)$$

where the numerical factor $\Gamma_{em}$ depends on the loop's shape, but not on its length; typically, $\Gamma_{em} \sim 100$. The ratio of the power in electromagnetic waves to that in gravitational waves is

$$\frac{\dot{E}_{em}}{\dot{E}_g} \sim \frac{e^2}{\hbar c}\left(\frac{G_m}{c^2}\right)^{-1}\left(\frac{J}{J_{max}}\right)^2, \qquad (45)$$

and we see that for sufficiently large electromagnetic radiation can become the dominant energy loss mechanism for the loop [11].

For a loop with kinks, $\Gamma_{em}$ in (44) has a weak logarithmic dependence on the current; its characteristic range is $10^2 \leq \Gamma_{em} \leq 10^3$. If the loop has cusps, then for $J \ll J_{max}$ the radiation power is dominated by the emission of short periodic bursts of highly directed energy from near-cups regions. An estimate of $\dot{E}$ in this case is complicated by the fact that the string motion near the cups is strongly affected by radiation back-reaction. Only an upper bound on this radiation power has been obtained,

$$\dot{E}_{em} \leq JJ_{max}/c. \qquad (46).$$

## 4. The Evolution of the 'New Formula'

### 4.1 Theoretical Approach

The Einstein's Famous Formula $\left(E = mc^2 = m_0 c^2 / \sqrt{1-\boldsymbol{b}^2}\right)$ is a general result of the Special Theory of Relativity (STR). Here, $m_0$ denotes the rest mass, $m$ denotes the relativistic mass and $\boldsymbol{b} = v/c$. According to the Einstein's Famous Formula (EFF) the energy ($E$) approaches infinity as the velocity ($v$) approaches the velocity of light ($c$). The velocity must therefore always remain less than $c$, however great may be energies used to produce the acceleration [1]. This means that, according to the EFF it is impossible for a particle to travel faster than light, and it is therefore impossible to escape from black holes.

As given in section (3.3.2) in detail; When an electric field is applied, the Fermi momentum grows as $\dot{p}_F = eE$, and the number of fermions per unit length, $n = p_F/2\boldsymbol{p}\hbar$, also grows [11]:

$$\dot{n} \sim eE/\hbar.$$

The particles move along the string at the speed of light. The resulting current is $J = enc$, and $dJ/dt$ is given by $dJ/dt = (ce^2/\hbar)E$.

The current continues to grow until it reaches a critical value

$$J_c \sim emc^2/\hbar,$$

when $p_F = mc$. At this point, particles at the Fermi level have sufficient energy to leave the string. Consequently, in this simplified picture, the growth of the current terminates at $J_c$ and the string starts producing particles at the rate (39). The fermion mass $m$ is model-dependent, but it does not exceed the symmetry breaking scale of the string, $\boldsymbol{h}$. Hence,

$$J_c \leq J_{max} \sim e(\boldsymbol{m}c^3/\hbar)^{1/2},$$

where we have used the relation $\boldsymbol{m} = \boldsymbol{h}^2 c/\hbar$.

This means that, the speed of particles approaches the speed of light $(v \rightarrow c)$ when the current inside string approaches the critical value $(J \rightarrow J_c = emc^2/\hbar)$. At this point, particles at the Fermi level can leave the string. Considering this result, it is quite clear that there is a direct relation between the speed of particles and the current value inside string. When we look the Einstein's Famous Formula (EFF), there is no the relation between the current value of mediums at which the particles move and the speed of particles. Thus, we can say that the Special Theory of Relativity (STR) and string theory are not compatible. Therefore, this paper describes a 'New Formula' by enhancing the EFF to provide the compatibility between STR and strings.

In addition, as given in sections 3.1 and 3.2, it is shown that rotating black holes produce electron-positron outflow when brought into contact with a strong magnetic field. It is quite clear that the current value inside black holes increases when they brought into contact with a strong magnetic field. This means that, the particles can escape from black holes when the current inside black holes approaches a critical value ($J \to J_c \leq J_{max}$). Thus, it is possible for particles to escape from black holes and consequently to reach to the speed of light $(v = c)$ which is not allowed by the Special Theory of Relativity. Therefore, It should be allowed by 'New Formula' that particles can reach and exceed the speed of light.

Considering the above requirements a 'New Formula' has been developed by enhancing the EFF to provide the compatibility between the Special Theory of Relativity (STR), black holes and strings. Here, $J/J_{max}$ has been added to the EFF as a new parameter, and the 'New Formula' has been developed as

$$E = \frac{m_0 c^2}{\sqrt{1 - \frac{v^2}{c^2}\left(1 - \frac{J}{J_{max}}\right)}} \ . \tag{47}$$

In case $J = J_c = emc^2/\hbar$ and $J_{max} = e(mc^3/\hbar)^{1/2}$ the 'New Formula' can also be written as

$$E = \frac{m_0 c^2}{\sqrt{1 - \frac{v^2}{c^2}\left(1 - \frac{emc^2/\hbar}{e(mc^3/\hbar)^{1/2}}\right)}} \ . \tag{48}$$

As the loop radiates away its energy by emitting electromagnetic and gravitational waves, it shrinks and the dc current in the loop grows as

$$J \propto L^{-1}. \tag{49}$$

If the loop has no cusps or kinks, then its electromagnetic radiation power is

$$\dot{E} \sim \Gamma_{em} J^2, \tag{50}$$

with $\Gamma_{em} \sim 100$. (The power is not much different for a kinky loop, but the presence of cusps can change it drastically.) The ratio of electromagnetic and gravitational power output is of the order

$$\frac{\dot{E}_{em}}{\dot{E}_g} \sim \frac{J^2}{Gm^2}. \tag{51}$$

If the current evolves according (49), then $\dot{E}_{em}$ gradually grows, and the net fraction of the loop's mass radiated electromagnetically during its entire lifetime is (Ostriker, Thompson and Witten, 1986)

$$f = \sqrt{f_i} \tan^{-1}\left(1/\sqrt{f_i}\right), \tag{52}$$

where $f_i$ is the initial value of $\dot{E}_{em}/\dot{E}_g$. Eventually the current reaches the critical level $J \approx J_{max}$. As the loop shrinks further, the current remains near critical and all the extra charge carriers are expelled from the strings [11].

For the loop with cusps, the electromagnetic radiation power is dominated by bursts of radiation from near-cusp regions. The motion of the string in these regions is strongly affected by the radiation back-reaction, and the resulting power is difficult to estimate. It is expected to be much greater than the power for a cuspless loop (50). An upper bound for $\dot{E}_{em}$ is given by

$$\dot{E}_{em} \leq J\sqrt{m}. \tag{53}$$

In the vicinity of a cusp, the current tends to become super-critical. An invariant measure of the current is

$$J_a J^a = q^2 (-g)^{-1/2} \left(f'^2 - \dot{f}^2\right). \tag{54}$$

Near a cusp $\sqrt{-g} = x'^2 \to 0$ and $J_a J^a \to \infty$ (unless $\dot{f} \pm f' = 0$). The physical origin of this effect is very simple: a moving string becomes contracted by a factor $|x'|^{-1} = (1 - \dot{x}^2)^{-1/2}$, and the density of charge carriers increases by the same factor. The right-hand side of (54) can be estimated, $J_a J^a \sim (JL/z)^2$. The current becomes super-critical in the region $|z| \leq LJ/J_{max}$. As a result the loop will lose a fraction $\sim J/J_{max}$ of its charge carriers. If the motion of the loop were strictly periodic, then during the next period the current near the cusp would be exactly $J_{max}$ [11].

Considering the above information, it is quite clear that the current value $(J)$ in a superconducting medium (e.g. inside string and black hole) can reach to a super critical value $(J_{max})$. This means that $J$ can be equal to $J_{max}$ in the 'New Formula', and in this state of the 'New Formula' $(J = J_{max})$ will allow that particles can reach to and exceed the speed of light and consequently can leave black holes and strings.

**State:** $J = J_{max}$

$$\begin{Bmatrix} v = 0 \\ E = m_0 c^2 \\ m = m_0 \end{Bmatrix}, \begin{Bmatrix} v = c \\ E = m_0 c^2 \\ m = m_0 \end{Bmatrix} \text{ and } \begin{Bmatrix} v = \infty \\ E = m_0 c^2 \\ m = m_0 \end{Bmatrix}.$$

There is no the limitation for the speed of a particle in this state. In addition, the energy and mass of a particle do not change when speed changes in this state. This state can predict and describe the space-time singularities without the distribution of mass and energy.

### 4.2 Practical Approach (Experiments)

The EEFF (Enhanced Einstein's Famous Formula) which is completely same as the 'New Formula' has been experimentally proved and justified. The detailed information about the practical approach of the'New Formula' (or EEFF) including experiments are given in the References [13] (cond-mat/9909373) and [14] (gr-qc/9909077).

## 5. Conclusion

A 'New Formula' has been theoretically developed and described in this paper. The 'New Formula' has been developed in place of Einstein's Famous Formula (EFF) to provide the compatibility between the Special Theory of Relativity (STR), black holes and strings. The 'New Formula' can also predict and describes the space-time singularities without the distribution of mass and energy. It is allowed by the 'New Formula' that any particle can reach to and exceed the speed of light $(v \geq c)$.

A very important conclusion of this paper is that the EFF $(E = mc^2)$ is only valid and applicable in the vacuum (the mediums which have low current density: outside the string, outside black hole), but is not valid and applicable for inside string and inside black hole including space-time singularities. However, the 'New Formula' is valid and applicable in all mediums including inside string and inside black hole.